\documentclass[aip, amsmath,amssymb, reprint]{revtex4-2}

\usepackage{graphicx}
\usepackage{dcolumn}
\usepackage{bm}
\usepackage[utf8]{inputenc}
\usepackage[T1]{fontenc}
\usepackage{newtxtext,newtxmath}
\usepackage{microtype}
\usepackage{amsmath}
\usepackage{mathrsfs}
\usepackage{bbm}
\usepackage{bm}
\usepackage{mathtools}
\usepackage{dsfont}
\usepackage{braket}
\usepackage{cancel}
\usepackage{slashed}
\usepackage{upgreek}
\usepackage{etoolbox}
\usepackage[separate-uncertainty = false]{siunitx}
\DeclareSIUnit\Gal{Gal}
\usepackage{multirow}
\usepackage{ifthen}
\usepackage[svgnames]{xcolor}

\definecolor{cset-aps-blueberry}{RGB}{28,128,158}
\definecolor{cset-aps-blue}{RGB}{46,44,184}
\definecolor{cset-aps-turquoise}{RGB}{0,67,88}
\definecolor{cset-aps-limegreen}{RGB}{190,219,67}
\definecolor{cset-aps-green}{RGB}{31,138,112}
\definecolor{cset-aps-yellow}{RGB}{255,225,25}
\definecolor{cset-aps-orange}{RGB}{253,116,0}
\definecolor{cset-aps-red}{RGB}{219,0,43}
\definecolor{sr-trans}{RGB}{207,0,0}
\definecolor{sr-ground}{RGB}{178,202,0}
\definecolor{sr-excited}{RGB}{11,70,135}
\definecolor{sr-mean}{RGB}{0,166,0}
\definecolor{sr-bs}{RGB}{207,0,255}

\definecolor{blau}{HTML}{1575B9}
\definecolor{hellblau}{HTML}{65B7EF}
\definecolor{rot}{HTML}{E31B0A}
\definecolor{hellrot}{HTML}{FC6761}
\definecolor{gruen}{HTML}{25A131}
\definecolor{lila}{HTML}{2E2CB8}
\definecolor{grau}{HTML}{A6A6A6}

\usepackage{layouts}

\usepackage{hyperref}
\hypersetup{
    colorlinks=true,
    linkcolor={cset-aps-red},
    linkbordercolor={cset-aps-red},
    filecolor={cset-aps-orange},
    filebordercolor={cset-aps-orange},
    citecolor={cset-aps-blue},
    citebordercolor={cset-aps-blue},
    urlcolor={cset-aps-green},
    urlbordercolor={cset-aps-green},
    menucolor={cset-aps-limegreen},
    menubordercolor={cset-aps-limegreen},
    breaklinks=true,
    pdfborderstyle={/S/U/W 2},
    pdfpagemode=UseOutlines,
    pdfstartpage={1},
}

\DeclareSIUnit{\au}{\text{a.u.}}
\DeclareSIUnit{\counts}{\text{counts}}
\DeclareSIUnit{\countunit}{\text{count}}
\DeclareSIUnit{\pipi}{\ensuremath{\pi}}

\newcommand{\ie}{i.\,e., }

\newcommand{\affTUDaSG}{
\affiliation{\href{https://ror.org/05n911h24}{Technische Universit{\"a}t Darmstadt}, Fachbereich Physik, Institut f{\"u}r Angewandte Physik, Schlossgartenstr. 7, 64289 Darmstadt, Germany
}
}

\newcommand{\affTUDaL}{
\affiliation{
    \href{https://ror.org/05n911h24}{Technische Universit{\"a}t Darmstadt},
    Fachbereich Physik,
    Institut f{\"u}r Angewandte Physik,
    Otto-Berndt-Str. 3,
    64287 Darmstadt,
    Germany
}
}

\newcommand{\orcid}[1]{\href{https://orcid.org/#1}{\includegraphics[width=7pt]{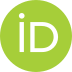}}}
\begin{document}

\title{
Phase estimation in spontaneous nonlinear interferometry for enhanced quantum imaging 
}

\author{Jonas L. Moos\,\orcid{0009-0004-6624-8380}}
\affTUDaL

\author{Cristofero Oglialoro\,\orcid{0009-0009-3162-1583}}
\affTUDaSG

\author{Enno Giese\,\orcid{0000-0002-1126-6352}}
\affTUDaSG

\author{Markus Gräfe\,\orcid{0000-0001-8361-892X}}
\affTUDaL


\begin{abstract}
Bicolor quantum imaging can reach, in principle, phase supersensitivity approaching the Heisenberg limit using squeezed light from high-gain parametric down-conversion, but established quantum-imaging setups typically operate in the low-gain, spontaneous regime, where such scaling is inaccessible.
Here, we show experimentally and theoretically that a symmetric nonlinear interferometer, even in the spontaneous regime, retains a phase-sensitivity advantage over configurations where entanglement is not exploited as a quantum-metrological resource, although the achievable phase sensitivity remains shot-noise-limited.
This advantage appears as a shift of the optimal working point toward the dark fringe, the low-gain signature of the mechanism underlying high-gain supersensitivity.
We derive design choices for phase-optimized nonlinear interferometers accordingly, favoring configurations that exploit entanglement as a metrological resource and specifying their optimal operating point.

\vspace{1mm}
\end{abstract}

\maketitle
\section{Introduction}
Quantum imaging~\cite{gilaberte_basset_perspectives_2019} allows to probe the optical properties of a sample with photons at one wavelength, while entangled photons at a conjugate wavelength are detected~\cite{lemos_quantum_2014, zou_induced_1991}.
In addition to such bicolor imaging with undetected photons, entanglement serves as a resource for quantum metrology and can enable phase sensitivities that scale inversely with the photon number, surpassing shot noise and reaching the Heisenberg limit~\cite{caves_quantum-mechanical_1981, giovannetti_advances_2011}.
While the benefits of such a Heisenberg scaling increase with higher photon numbers~\cite{giovannetti_advances_2011}, this work demonstrates experimentally and theoretically that a phase-sensitivity advantage in quantum imaging persists even in the spontaneous regime of photon-pair generation, although the achievable phase sensitivity remains shot-noise-limited.
    
\begin{figure}
    \centering
    \includegraphics[width= \linewidth]{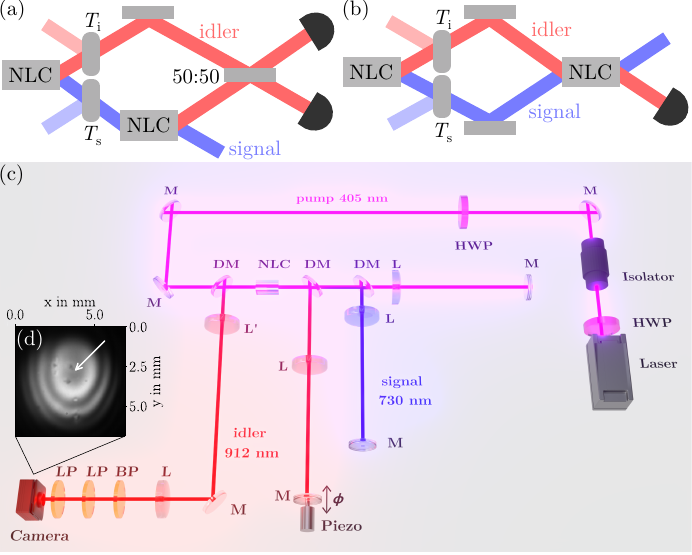}
    \caption{
    Nonlinear interferometers for quantum imaging and experimental realization.
    (a) In the induced coherence (IC) setup, a nonlinear crystal generates signal and idler fields (blue and red).
    The signal field seeds a second nonlinear crystal, and the two idler modes are subsequently interfered on a beam splitter before detection.
    (b) In an SU(1,1) interferometer, both the signal and idler fields seed the second nonlinear crystal, and the resulting idler field is detected at its output.
    (c) Experimental SU(1,1) setup in a retroreflective implementation used in this work, including the pump preparation stage (purple). The phase $\phi$ of the idler beam (red) is controlled by a piezo stage before detection.
    Components: dichroic mirrors (DM), lenses with focal lengths of $\qty{150}{\milli \meter}$ (L) and $\qty{100}{\milli \meter}$ (L'), mirrors (M), band-pass filter (BP), long-pass filter (LP), half-wave plate (HWP), nonlinear crystal (NLC).
    (d) Representative camera image at an arbitrary phase.
    The arrow indicates the pixel used for evaluation.
    }
    \label{fig:setup}
\end{figure}

Two nonlinear interferometer configurations are commonly used for bicolor quantum imaging~\cite{Miller2021_Versatileb,Kranias2025_Metrologicala,fuenzalida2024c}:
The induced-coherence (IC) configuration~\cite{zou_induced_1991,wang_induced_1991}, see \autoref{fig:setup}\,(a), has an intrinsic phase sensitivity that scales as $1/\sqrt{n}$ and is therefore shot-noise-limited in the number $n$ of generated photons~\cite{kolobov_controlling_2017,Wiseman2000_Induceda}.
The symmetric SU(1,1) configuration~\cite{yurke_su2_1986, herzog_frustrated_1994}, see \autoref{fig:setup}\,(b), offers improved robustness~\cite{gilaberte_basset_videorate_2021} and, in principle, reaches a Heisenberg-limited phase-uncertainty scaling~\cite{manceau_improving_2017, giese_phase_2017, florez2018a, schaffrath2024c} as $1/\sqrt{n(n+1)}$.
Especially in the high-parametric-gain regime~\cite{Frascella_19_Wide-field,Lindner2023_High-sensitivity,Machado2020_Optical} with $1\ll n$, the different phase-sensitivity scaling of the two configurations leads to a clear advantage of the SU(1,1) interferometer.
However, this is largely restricted to the idealized, loss-free case.
In the presence of loss, the phase uncertainty of the SU(1,1) interferometer degrades~\cite{marino_effect_2012} to shot-noise scaling or even a constant noise floor~\cite{oglialoro_below-shot-noise_2026}, unless the interferometer is deliberately gain-unbalanced~\cite{manceau_detection_2017,manceau_improving_2017, giese_phase_2017}.
    
The extreme susceptibility of supersensitivity to loss stems from the underlying working principle:
Ideally, two-mode squeezed states are generated, acquire phase, and are anti-squeezed back to the vacuum state in frustrated down-conversion~\cite{herzog_frustrated_1994}.
Hence, SU(1,1) interferometers are usually operated at destructive interference (dark fringe) at phase $\phi = \pi$.
Loss deteriorates the quantum state so that perfect frustrated down-conversion is not possible.
Instead, the optimal phase bifurcates and shifts towards mid-fringe detection for increasing loss~\cite{marino_effect_2012,xin2019}, making true phase supersensitivity challenging to observe~\cite{Du2018_Absolutea,Linnemann2016_QuantumEnhancedb,manceau_detection_2017} and achievable only under ideal conditions or with homodyne detection~\cite{Li2019_Pulsed}.

While Heisenberg scaling is most compelling in the high-parametric-gain regime~\cite{ou_quantum_2020}, it evades observation since quantum imaging aimed at real-world applications is primarily performed in the spontaneous, low-parametric-gain regime~\cite{chekhova_nonlinear_2016,kutas2020b,fuenzalida2024c}.
In this regime~\cite{Santandrea2023_Lossy,Lindner_21_Nonlinear,Gattinger2026} at $n\ll1$, both setups lead to a shot-noise-limited phase uncertainty~\cite{kolobov_controlling_2017, manceau_detection_2017} $1/\sqrt{n}$, so there is no obvious benefit besides the inherent symmetry and robustness of the SU(1,1) configuration.
This work demonstrates, for the first time, that the phase-sensitivity advantage observed in the high-parametric-gain regime extends to spontaneous photon-pair generation, despite the absence of Heisenberg-limited scaling.

\section{Setup}

\autoref{fig:setup}\,(c) shows the nonlinear interferometer in the SU(1,1) configuration used in this work. 
A laser, protected against back-reflections by an optical isolator, emits a pump beam at \qty{405}{\nano \meter} with a power of \qty{57(3)}{\milli \watt}.
The polarization of the pump is set by a half-wave plate to satisfy the phase-matching condition of a nonlinear periodically poled potassium titanyl phosphate (PPKTP) crystal with a length of \qty{1}{\milli \meter}, temperature-controlled at \qty{29.34}{\celsius}.
Inside the crystal, photon pairs at an idler wavelength of \qty{912}{\nano \meter} and a signal wavelength of \qty{730}{\nano \meter} are generated by type-0 spontaneous parametric down-conversion.
The spontaneous nature of the process was confirmed separately by recording the idler intensity $I$ as a function of the pump power and observing a linear dependence (see Appendix~\ref{sec:technoise}). In this spontaneous regime, the signal and idler intensities are expected to exhibit Poissonian statistics and shot-noise scaling, namely $\Delta I^2 \propto I$ (confirmed in \autoref{sec:results}). 

After the crystal, the pump, signal, and idler beams are spatially separated by dichroic mirrors, and the crystal plane is imaged onto itself for each wavelength by a $4f$ system consisting of a lens with \qty{150}{\milli \meter} focal length and a mirror. 
The mirror in the idler arm is mounted on a piezo-translation stage, enabling a scan of the phase. 
On the second pass through the crystal, the idler beam is separated again from the pump and signal beams by a dichroic mirror and magnified by a factor of $\num{1.5}$ in a $4f$ system of two lenses with \qty{100}{\milli \meter} and \qty{150}{\milli \meter} focal length, respectively.
Two long-pass filters, each with a cut-on wavelength of \qty{500}{\nano \meter}, along with a band-pass filter centered at \qty{910}{\nano \meter} with a bandwidth of \qty{10}{\nano \meter} corresponding to the full width at half maximum, suppress background light.
The idler is then detected by a qCMOS camera (Hamamatsu ORCA-Quest 2).

For the measurement, the phase $\phi = \theta + l \cdot 2 \pi   \delta_\mathrm{s}/\lambda_\mathrm{fr}$ is varied by scanning the piezo stage over \qty{1}{\micro \meter} in steps of $\delta_\mathrm{s} =\qty{2}{\nano \meter}$. At each phase position $l=1,\dots,\num{500}$, we record $m=1,\dots,\num{30}$ intensity images with an exposure time of \qty{1}{\second} [see the representative image in \autoref{fig:setup}\,(d)].
Here, $\theta$ denotes the offset phase of the interference signal relative to the initial phase and $\lambda_\mathrm{fr}$ is the fringe spacing, both extracted from the fit depicted in \autoref{fig:measurements}\,(b) and described below.
For the evaluation, only the intensity of the pixel marked by the arrow in \autoref{fig:setup}\,(d) is used, yielding the sets of intensities $I_m(\phi)$. This pixel is representative of all pixels that show no obvious distortion of the interference pattern (see Appendix~\ref{sec:neighborhood}).

\section{Results} \label{sec:results}
    \begin{figure}
        \centering
        \includegraphics[]{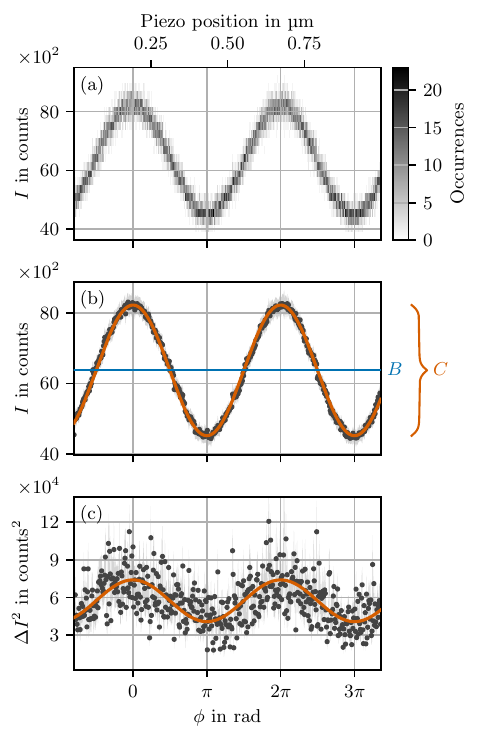}
        \caption{
        Phase-resolved measurement of the idler intensity of a single pixel. (a) Photon statistics shown by histograms of \num{30} measured intensities, sorted into \num{20} bins of \qty{267.15}{\counts} width for each phase setting.
        (b) Fringe scan obtained from the mean intensity $I$ with its standard deviation (shaded region); the red line is a fit of Eq.~\eqref{eq:interference}.
        (c) Intensity variance $\Delta I^2(\phi)$ with its uncertainty (shaded region), showing the phase-dependent (heteroscedastic) fluctuations; the red line is the fringe fit from (b) scaled by $\alpha = \qty{9.0(1)}{\counts}$.
        }
        \label{fig:measurements}
    \end{figure}
For each phase setting, a histogram is constructed from this measurement, in which all recorded intensities are sorted into \num{20} equally spaced intensity bins of \qty{267.15}{\counts} width, see \autoref{fig:measurements}\,(a). 
The whole distribution of recorded intensities follows the interference fringe.
    
At each phase setting, the mean intensity $I(\phi)=\braket{I_m(\phi)}$ is calculated and shown in \autoref{fig:measurements}\,(b), with the associated standard deviation represented as a shaded region.
The observed interference pattern of an SU(1,1) interferometer can be described as
\begin{align}\label{eq:interference}
    I(\phi)=B (1+ C \cos \phi)\text{\,.}
\end{align}
From the fit displayed as a red line, the baseline $B=\qty{6370(4)}{\counts}$, the contrast $C=\num{0.2899(9)}$, the fringe spacing $\lambda_\mathrm{fr}=\qty{480.7(4)}{\nano \meter}$, and the offset phase $\theta=\qty{-2.508(6)}{\radian}$ are obtained; the latter two define the interferometer phase introduced above. 
The fringe spacing corresponds approximately to half the idler wavelength; the deviation can be attributed to an inaccuracy in the calibration of the piezo element and to a constant phase drift over the duration of the measurement.
The stated uncertainties are the fit uncertainties.

The measured baseline can be related to the theoretical expression $B= \alpha n \eta (1+T_\mathrm{i})$ in the spontaneous regime~\cite{oglialoro_below-shot-noise_2026}. 
Here, $n$ is the number of photon pairs generated in the first pass through the crystal, $\alpha$ is the camera gain converting the detected photon number into counts, $\eta$ is the detection efficiency including external loss, and $T_\mathrm{i}$ is the transmittance of the idler arm shown in \autoref{fig:setup}\,(b), including internal losses and object properties. 
The corresponding contrast in the spontaneous regime~\cite{oglialoro_below-shot-noise_2026} is $C = 2 \sqrt{T_\mathrm{s}T_\mathrm{i}} / (1+T_\mathrm{i})$, where $T_\mathrm{s}$ is the transmittance of the signal arm. The contrast is independent of $n$ and $\eta$, assuming phase stability.
Hence, internal loss inevitably reduces the contrast. While near-unit contrast is crucial for quantum-enhanced phase sensitivity, achieving it is the most difficult condition to meet experimentally, since it requires near-perfect anti-squeezing, which is extremely susceptible to internal loss.

\begin{figure}
        \centering
        \includegraphics[]{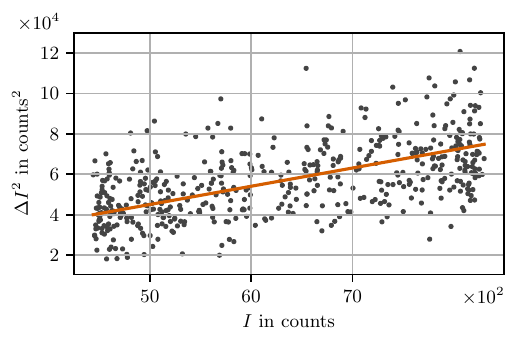}
        \caption{
        Scatter plot of the intensity fluctuations $ \Delta I^2$ versus the corresponding mean intensity $I$ for all phase settings of the phase scan. 
        The linear dependence (red line) demonstrates the Poisson statistics expected in the spontaneous regime, with slope $\alpha=\qty{9.0(1)}{\counts}$.
        }
        \label{fig:technoise}
    \end{figure}

\autoref{fig:measurements}\,(c) shows the intensity fluctuations at each phase, calculated as the sample variance $\Delta I^2(\phi)= \braket{I_m^2(\phi)}-I^2$ of the intensity statistics at each phase setting. Their uncertainties are obtained from the variance of the sample variance using the fourth moment~\cite{cho_variance_2008} and are displayed as a shaded region.
The fluctuations are heteroscedastic~\cite{hudelist_quantum_2014}, \ie they vary with the phase, indicating that they cannot be explained by phase-independent baseline noise alone.
In fact, the fluctuations follow the same interference pattern as the intensity: the red line in \autoref{fig:measurements}\,(c) is the intensity fit from \autoref{fig:measurements}\,(b), rescaled by the conversion gain of the camera $\alpha$.
As described in the definition of $B$, the conversion gain $\alpha$ relates the detected intensity to the number of incident photons.
Since the intensity scales linearly with $\alpha$ while its variance scales quadratically ($\propto\alpha^2$), the measured ratio is $\Delta I^2/I = \alpha$ for Poissonian photon statistics.
To determine $\alpha$, we plot the fluctuations $\Delta I^2$ extracted from the data of \autoref{fig:measurements} against $I$ as a scatter plot in \autoref{fig:technoise}. 
They follow the linear dependence, indicating shot-noise scaling, since the thermal statistics of down-conversion reduce to Poissonian statistics in the spontaneous regime. 
A linear fit (red line, assuming negligible offset at $I=0$) yields the conversion gain $\alpha = \Delta I^2/I = \qty{9.0(1)}{\counts}$. 
This agrees with the independent measurement in Appendix~\ref{sec:technoise}, where the influence of the detection scheme, \ie the camera, on $\alpha$ is characterized.

The key to achieving Heisenberg-scaling high-gain in SU(1,1) interferometry is to work at the point of destructive interference, where the intensity vanishes and the fluctuations are therefore suppressed.
However, this is possible only for perfect contrast~\cite{marino_effect_2012}.
Similar to the high-gain regime, high-precision phase estimation in the spontaneous regime with internal losses does not require minimizing the photon-number fluctuations, but the phase uncertainty.
Like the interference signal and the intensity fluctuations, the phase uncertainty is heteroscedastic and depends on the phase.
To extract the estimated phase information $\Phi$ at each phase setting $\phi$ from the measured intensities, the Eq.~\eqref{eq:interference} is inverted for each of the \num{30} measurements, so that, with the fitted values of $B$ and $C$ obtained from the mean intensity, the phase can be retrieved from a single image as
\begin{align} \label{eq:phase_retr}
    \Phi_{m} = \mp \arccos\Bigg(\frac{I_m(\phi) - B}{B C}\Bigg) + k\cdot 2\pi \text{\,,}
\end{align}
where $k = 0,1,2$ accounts for the periodic ambiguity of the $\arccos$.
The combined phase estimate is obtained as the average of these single-image estimators, $\Phi=\braket{\Phi_{m}}$.
However, due to the nonlinear intensity-to-phase conversion, the estimator is generally biased~\cite{Box1971_Bias}, resulting in a systematic error $\Delta\Phi_\mathrm{sys} =\Phi -\phi $.
This error can be reduced using more advanced estimation approaches~\cite{Pezze2018_Quantumm,derr_parameter_2026}. The statistical uncertainty, on the other hand, is given by $\Delta \Phi^2_\mathrm{stat}= \braket{(\Phi_{m}-\Phi)^2}$.
Near mid-fringe, $\Delta\Phi_\mathrm{sys}^2$ remains below \qty{4e-6}{\radian} and
contributes less than \qty{0.1}{\percent} of $\Delta\Phi^2$.

\begin{figure}
        \centering
        \includegraphics[]{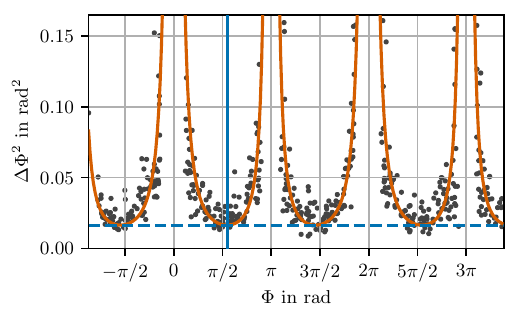}
        \caption{
        Phase sensitivity and optimal working point.
        Phase-estimation variance as a function of the reconstructed phase (black dots).
        The theoretical prediction (red line) reaches its minimum at $\Phi_\mathrm{min}$ (solid blue line), which lies between the IC benchmark at $\pi/2$ and the ideal dark-fringe working point at $\pi$; the corresponding uncertainty $\Delta \Phi_\mathrm{min}^2$ is indicated by the dashed blue line.
        }
        \label{fig:phaseuncertainty}
    \end{figure}

\autoref{fig:phaseuncertainty} displays the phase uncertainty $\Delta \Phi^2= \Delta \Phi^2_\mathrm{stat}+\Delta\Phi_\mathrm{sys}^2$ as a function of the phase.
The theoretical description relies on Gaussian uncertainty propagation
\begin{align}
    \Delta \Phi^2 = \frac{\Delta I^2}{|\partial_\Phi I|^2} = \frac{\alpha (1+C \cos \Phi)}{B C^2 (1 - \cos^2 \Phi)}
\end{align}
in the spontaneous regime, which is computed with the values for $B$ and $C$ determined by the fits in \autoref{fig:measurements}\,(b) and for $\alpha$ determined in \autoref{fig:technoise}.
The result is displayed in \autoref{fig:phaseuncertainty} as a red line.

The solid vertical blue line, marking the phase of minimal phase uncertainty, deviates from the position of destructive interference at $\Phi = \pi$ as a result of the interplay between small photon-number fluctuations and the slope of the interference pattern.
The optimal working point of the interferometer for phase measurements in the spontaneous regime is therefore given by~\cite{oglialoro_below-shot-noise_2026}
\begin{align}\label{eq:phimin}
    \Phi_\mathrm{min}= \arccos[ \sqrt{C^{-2}-1}- C^{-1}] \text{\,,}
\end{align}
which depends only on the contrast and approaches $\pi$ for perfect contrast.
The minimal phase uncertainty obtained from
\begin{align}\label{eq:deltaphimin}
    \Delta \Phi_\mathrm{min}^2 = \frac{\alpha}{2 B  [1-\sqrt{1-C^2}]}\text{\,,}
\end{align}
shown as a dashed blue line in \autoref{fig:phaseuncertainty}. 
The theoretical predictions of $\Phi_\mathrm{min}= \qty{1.7195(5)}{\radian}$ and $\Delta \Phi_\mathrm{min}^2 = \qty{0.0165(2)}{\radian^2}$, calculated with Eqs.~\eqref{eq:phimin} and \eqref{eq:deltaphimin} from the fit parameters in \autoref{fig:measurements}\,(b), respectively, are in excellent agreement with the experimental data.

\section{Discussion}
In the spontaneous regime, the SU(1,1) interferometer reaches shot-noise scaling with $B\sim n$.
Although the setup operates at a camera conversion gain $\alpha > \qty{1}{\countunit}$, the phase uncertainty exhibits the same shot-noise scaling with $n$ as in the ideal case.
Heisenberg scaling, in contrast, is expected only to be apparent in the high-parametric-gain regime~\cite{ou_quantum_2020,chekhova_nonlinear_2016}.

Nevertheless, the benefit of the SU(1,1) setup over the IC interferometer of \autoref{fig:setup}\,(a) persists in the spontaneous regime. 
In a single-exit-port detection scheme, where only one of the two outputs of the beam splitter is detected, the IC signal has the same form as Eq.~\eqref{eq:interference}.
In the spontaneous regime, it also exhibits the same contrast as the SU(1,1) configuration, which leads to the same optimal working point.
Only the baseline is reduced by a factor of two to $B/2$ for an ideal 50/50 beam splitter.
Hence, assuming identical loss and noise statistics, an IC interferometer results in a factor-of-two increase in the phase-uncertainty variance  $\Delta \phi_\mathrm{min}^2$.

However, the single-exit-port scheme is not the optimal mode of operation, as light from the other exit port of the beam splitter is discarded.
Differential operation, where the intensity difference between the two exit ports of an IC interferometer is measured, is preferred~\cite{Miller2021_Versatileb,Kranias2025_Metrologicala}
and yields $I_-(\phi)= B (1+C\cos\phi)/2-B (1-C\cos\phi)/2= B C \cos \phi$.
Such a differential detection scheme could, in principle, suppress additional technical noise.
For a fair comparison with the SU(1,1) configuration, however, we assume the same conversion gain $\alpha$ for both schemes, which gives rise to the phase uncertainty
\begin{align}
    \Delta \phi_-^2 =  \frac{\alpha + B C^2 \cos^2 \phi}{B C^2 (1-\cos^2\phi)}
\end{align}
with shot-noise behavior, because $B\sim n$. This gives a minimum at mid-fringe ($\phi = \pi/2$), where $\Delta \phi_-^2(\pi/2) =\alpha/(BC^2)$, which is the value expected for a conventional shot-noise-limited interferometer geometry.
With this assumption of the same conversion gain and technical noise, the differential IC detection scheme and the SU(1,1) configuration agree at mid-fringe, $\Delta \Phi^2(\pi/2) = \Delta \phi_-^2(\pi/2)$.

Since the noise behavior of such a setup can only be characterized unambiguously in a hybrid setup~\cite{Gemmell2024_Couplinga}, the mid-fringe value of this work is used as a benchmark in the following.
The optimal working point of the SU(1,1) interferometer is not at mid-fringe, but at $\Phi_\mathrm{min} = \qty{1.7195(5)}{\radian} = \qty{0.54733(2)}{\pipi}$ with a slight, but detectable, deviation from mid-fringe.
This results in an improvement by a factor of
\begin{align}
    \frac{\Delta \Phi^2(\pi/2)}{\Delta\Phi_\mathrm{min}^2} = \frac{2}{1+\sqrt{1-C^2}} = \num{1.0220(1)}
\end{align}
for the present contrast.
In the limit of perfect contrast ($C\rightarrow1$), this improvement grows to a factor of two.

In summary, the phase sensitivity of an SU(1,1) interferometer was analyzed in the spontaneous regime~\cite{Santandrea2023_Lossy}.
This work demonstrates that the working point of minimal phase uncertainty is not at mid-fringe but approaches the dark fringe for perfect contrast, in line with earlier predictions for the lossy high-gain regime~\cite{xin2019}.
In this regime, supersensitive measurements are possible.
In contrast, in shot-noise-limited interferometers such as IC or classical interferometers, the point of optimal operation is expected to be at mid-fringe when a differential detection scheme is employed. 
However, such a detection cannot be realized in bicolor imaging with an SU(1,1) interferometer, where recombination is nonlinear rather than through a beam splitter and only one wavelength is detected.

Theoretically, in the spontaneous regime, the SU(1,1) interferometer outperforms the IC configuration in the single-exit-port detection mode by a factor of two in its phase-uncertainty variance, assuming the same noise statistics.
Extrapolating from the mid-fringe benchmark, even the differential-exit-port mode of an IC interferometer remains inferior in the absence of technical noise.
Nevertheless, a direct comparison requires a hybrid or combined setup~\cite{Gemmell2024_Couplinga,oglialoro_below-shot-noise_2026} with equal loss conditions in both modes, in which primarily the technical noise of the IC differential mode would need to be studied.

Therefore, the SU(1,1) interferometer exhibits a phase-sensitivity advantage in the spontaneous regime that is the low-gain counterpart of the quantum-enhanced phase  supersensitivity achieved in the high-parametric-gain regime.
As photon-pair numbers grow exponentially, technical noise potentially loses its impact relative to the spontaneous regime.
However, the high-parametric-gain regime relies on a highly entangled state, which makes it more susceptible to loss~\cite{marino_effect_2012} and complicates the operation of an SU(1,1) interferometer, especially in quantum imaging setups with loss.

Already in the low-gain regime, where most nonlinear interferometers and quantum imaging experiments are performed~\cite{Topfer:25_Synthetic,Fuenzalida2023_Experimental}, the SU(1,1) interferometer exhibits a phase-sensitivity advantage over the IC configuration.
This intrinsic benefit survives beyond the high-parametric-gain regime, where quantum-enhanced phase supersensitivity emerge.
The advantage increases with the contrast, approaching a factor of two in the limit of perfect contrast.


\section*{Acknowledgments}
This work was funded by the Deutsche Forschungsgemeinschaft (DFG, German Research Foundation) – project number 552245798. Further support is acknowledged from the German Federal Ministry of Research, Technology and Space (BMFTR, formerly BMBF) within the funding program ``quantum technologies -- from basic research to market'' with contract number 13N16496 (QUANCER). 
\noindent We thank D. Derr for helpful feedback.

\section*{Author Declarations}

\subsection*{Conflict of Interest}
\noindent The authors have no conflicts to disclose.

\subsection*{Author Contributions}
\noindent \textbf{Jonas L. Moos:} Methodology (lead); Software (equal); Validation (lead); Formal analysis (equal); Investigation (lead); Data Curation (lead); Writing -- original draft (lead); Writing -- review \& editing (equal); Visualization (lead).
\textbf{Cristofero Oglialoro:} Methodology (supporting); Conceptualization (lead); Software (equal); Formal analysis (equal); Investigation (supporting); Writing -- review \& editing (equal).
\textbf{Enno Giese:} Conceptualization (supporting); Methodology (supporting); Writing -- review \& editing (equal); Supervision (equal); Funding acquisition (equal); Project administration (equal).
\textbf{Markus Gräfe:} Methodology (supporting); Resources (lead); Writing -- review \& editing (equal); Supervision (equal); Funding acquisition (equal); Project administration (equal).

\section*{Data availability}
\noindent The data that support the findings of this study are available from the corresponding author upon reasonable request.

\appendix
\section{Conversion gain}\label{sec:technoise}
\begin{figure}[h]
    \centering
    \includegraphics[]{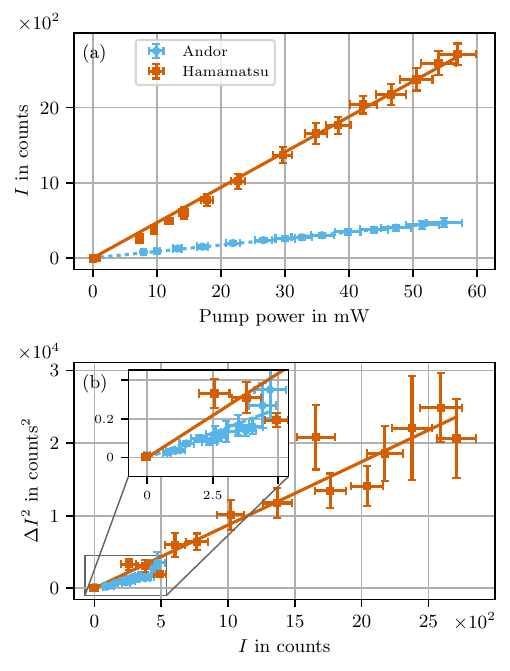}
    \caption{
    (a) Dark-image-subtracted idler intensity as a function of pump power for the Andor (blue) and the Hamamatsu (red) camera, with linear fits of negligible offset.
    The error bars of the pump power correspond to the \qty{5}{\percent} accuracy quoted by the manufacturer.
    (b) Scatter plot of the idler intensity variance versus the corresponding mean idler intensity, with a linear fit assuming negligible offset.
    }
    \label{fig:cameras}
\end{figure}
To verify the operation in the low-gain regime, the idler intensity is recorded for different pump powers and by that at different parametric gains.
Only idler photons generated in the first pass of the crystal are detected by blocking the pump and signal beam behind the crystal.
The intensities at one selected pixel are recorded with two different cameras (a Hamamatsu ORCA-Quest 2 qCMOS camera and an Andor iStar sCMOS) and averaged over \num{30} images.
\autoref{fig:cameras}\,(a) shows the expected linear increase of the intensity, confirming an operation in the spontaneous regime of down-conversion.
However, the slopes differ substantially between the cameras.

To analyze their conversion gain, we calculate the intensity variance $\Delta I ^2$ from these \num{30} images recorded in the single-pass experiment and plot them as a scatter plot against the corresponding mean intensity $I$ in \autoref{fig:cameras}\,(b).
While both show a linear dependence $\Delta I^2 = \alpha I$ with negligible offset, fits to the two datasets yield different slopes $\alpha = \qty{5.1(4)}{\counts}$ and $\alpha = \qty{8.7(4)}{\counts}$ for the two camera models Andor and Hamamatsu, respectively.
The conversion gain is therefore consistent with the value obtained from \autoref{fig:technoise}; the small remaining difference lies within the measurement uncertainty and is compatible with the different conditions of the single-pass characterization and the two-pass interferometric measurement.
In conclusion, the conversion gain $\alpha$ depends on the camera, while Poissonian statistics is observed in all cases.

\section{Neighborhood of the analysis pixel}
\label{sec:neighborhood}
\begin{figure*}
    \centering
    \includegraphics[width=\textwidth]{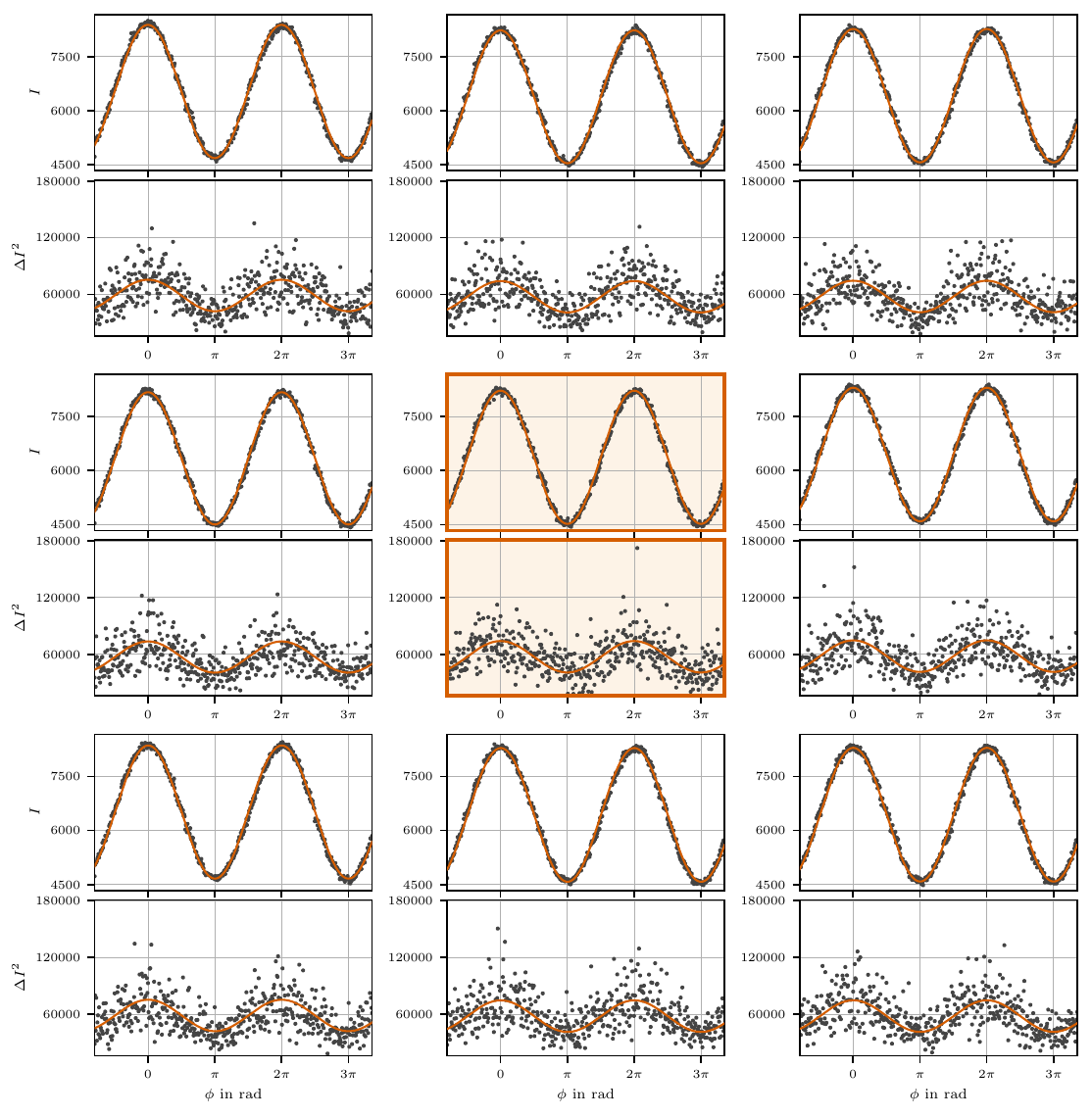}
    \caption{
    Intensity and variance across the $3\times3$ pixel neighborhood around the analysis pixel.
    Each cell corresponds to one sensor pixel, arranged by its position; the highlighted central cell is the analysis pixel used throughout the main article.
    Within each cell, the upper panel shows the mean intensity $I$ and the lower panel the variance $\Delta I^2$ as a function of the phase $\phi$.
    Points are the per-phase statistics over the \num{30} repetitions; the red line is the fit of Eq.~\eqref{eq:interference} to $I$, and in the variance panels the same fit scaled by the conversion gain, $\Delta I^2 = \alpha\,I$.
    }
    \label{fig:surrounding}
\end{figure*}

To confirm that the arbitrary choice of the analysis pixel does not bias the results, the intensity and variance analysis is repeated for the eight surrounding pixels, \ie the full $3\times3$ neighborhood.
\autoref{fig:surrounding} shows the mean intensity and variance of each pixel together with the fit of Eq.~\eqref{eq:interference}, arranged in a $3\times3$ matrix.
\autoref{tab:neighborhood_fit} lists the corresponding fit parameters.

All parameters agree across the neighborhood to within a few percent: the contrast $C$ and the baseline $B$ vary by at most $\sim\!\qty{3}{\percent}$, while the fringe spacing $\lambda_\mathrm{fr}$ and the offset phase $\theta$ are constant to well below \qty{1}{\percent}.
Since the optimal working point $\phi_\mathrm{min}$ depends only on the contrast, this spread translates into a negligible variation of the reported working point and phase sensitivity. The analysis pixel is therefore representative, and the results do not depend on its specific choice.
\begin{table}
  \centering
    \caption{
    Fit parameters of Eq.~\eqref{eq:interference} for the $3\times3$ neighborhood around the analysis pixel; the $(0,0)$ row is the analysis pixel used in the main text.
    }  
    \label{tab:neighborhood_fit}
  \begin{tabular}{l S[table-format=1.4(2)] S[table-format=3.1(1)] S[table-format=-1.3(2)] S[table-format=4(1)]}
    \hline\hline
    $(\Delta y,\Delta x)$ & {$C$} & {$\lambda_\mathrm{fr}$ (\si{\nano\meter})} & {$\theta$ (\si{\radian})} & {$B$ (\si{\counts})} \\
    \hline
    $(-1,-1)$ & 0.2829(9) & 480.5(4) & -2.507(6) & 6533(4) \\
    $(-1,\phantom{-}0)$ & 0.2886(9) & 480.5(4) & -2.513(6) & 6385(4) \\
    $(-1,+1)$ & 0.2883(9) & 480.3(4) & -2.516(6) & 6417(4) \\
    $(\phantom{-}0,-1)$ & 0.2893(9) & 480.6(4) & -2.512(6) & 6340(4) \\
    $(\phantom{-}0,\phantom{-}0)$ & 0.2899(9) & 480.7(4) & -2.508(6) & 6370(4) \\
    $(\phantom{-}0,+1)$ & 0.2874(9) & 480.5(4) & -2.518(6) & 6445(4) \\
    $(+1,-1)$ & 0.2835(9) & 480.2(4) & -2.523(6) & 6512(4) \\
    $(+1,\phantom{-}0)$ & 0.2884(9) & 480.5(4) & -2.521(6) & 6433(4) \\
    $(+1,+1)$ & 0.2880(9) & 480.5(4) & -2.521(6) & 6439(4) \\
    \hline\hline
  \end{tabular}
\end{table}

\section*{References}
\bibliography{literatureClean}

\end{document}